\documentclass[journal,comsoc]{IEEEtran}
\usepackage[T1]{fontenc}% optional T1 font encoding
\usepackage{color}
\usepackage{amsmath}
\usepackage{url}
\usepackage{array}
\usepackage{fancyhdr}
\usepackage{latexsym}
\usepackage{diagbox}
\usepackage{epsfig}
\usepackage{amssymb}
\usepackage{amsfonts}
\usepackage{amsxtra}
\usepackage{xspace}
\usepackage{makeidx}
\usepackage{graphics}
\usepackage{theorem}
\usepackage{subfigure}
\usepackage{multirow}
\usepackage{verbatim}
\usepackage{algorithmic,algorithm}
\DeclareMathOperator*{\minimize}{minimize}

\usepackage[noadjust]{cite}
\newtheorem{lemma}{\textbf{Lemma}}

\begin{document}

\title{Hybrid Beamforming Optimization for DOA Estimation Based on the CRB Analysis}

\author{Tian Lin, Xuemeng Zhou, Yu Zhu, and  Yi Jiang
\thanks{This work was supported by National Natural Science Foundation of China under Grant No. 61771147 and No. 61771005.}
\thanks{The authors are with the Key Laboratory for Information Science of Electromagnetic Waves (MoE), School of Information Science and Technology, Fudan University, Shanghai 200433, China. (e-mail: lint17@fudan.edu.cn, xuemeng19@fudan.edu.cn, zhuyu@fudan.edu.cn, jiangyi@fudan.edu.cn).}
}

\maketitle

\begin{abstract}
Direction-of-arrival (DOA) estimation is one of the most demanding tasks for the millimeter wave (mmWave) communication of massive multiple-input multiple-output (MIMO) systems with the hybrid beamforming (HBF) architecture. In this paper, we focus on the optimization of the HBF matrix for receiving pilots to enhance the DOA estimation performance. Motivated by the fact that many existing DOA estimation algorithms can achieve the  Cram\'{e}r-Rao bound (CRB), we formulate the HBF optimization problem aiming at minimizing the CRB with the prior knowledge of the rough DOA range. Then, to tackle the problem with intractable non-convex constraint introduced by the analog beamformers, we propose an efficient manifold optimization (MO) based algorithm. Simulation results demonstrate the significant improvement of the proposed CRB-MO algorithm over the conventional random HBF algorithm, and provide insights for the HBF design in the beam training stage for practical applications. 
%It is elaborated the optimized HBF for DOA measurements, can significantly outperform the traditional random receivers by sufficiently utilizing the prior knowledge of the rough DOA range (existing whether for warm boot or cold boot in practical). The optimization problem is formulated based on the Cram\'{e}r-Rao bound (CRB) analysis, and a manifold optimization based algorithm is proposed to efficiently solve the intractable problem with non-convex constraints resulting from the hardware limitation. Simulations show that dramatic improvements can be achieved, and provide the design insights of optimization of the training HBF, which is applicable for different specific DOA estimation algorithms. 
\end{abstract}

\begin{IEEEkeywords}
Hybrid beamforming, Cram\'{e}r-Rao bound, direction-of-arrival estimation, manifold optimization.
\end{IEEEkeywords}
\IEEEpeerreviewmaketitle

\section{Introduction}\label{sec:introduction}
%to guarantee enough signal-to-noise ratio (SNR) at the receiver side  for  . By separating the whole beamformer into a low-dimensional baseband digital one and a high-dimensional analog one implemented with phase shifters, the HBF architecture is enable to dramatically reduce the number of required radio frequency (RF) chains and achieve a good balance between performance and hardware cost \cite{lin_hybrid_2019}.
Hybrid beamforming (HBF) is regarded as a promising technology for millimeter wave (mmWave) massive multiple-input  multiple-output (MIMO) communication systems due to its advantage of achieving considerable beamforming gains with much lower hardware cost and power consumption when compared with the fully digital beamforming  \cite{yu_alternating_2016, ayach_spatially_2014}. However, its performance heavily relies on the accuracy of direction-of-arrival (DOA) estimation. There have been many works focusing on the design of DOA estimation algorithms with the HBF architecture. For example, an algorithm using the two-dimension (2D) discrete Fourier
transform (DFT) approach has been proposed in \cite{2018FanD}. Subsequently, a fast root multiple signal classification (root-MUSIC) algorithm has been developed in \cite{shu_low-complexity_2018} by extending the conventional MUSIC algorithm. 

For massive MIMO systems, the hybrid beamformers of high dimensions are endowed with sufficient freedom to customize the baseband pilots and benefit the subsequent DOA estimation. However, most related works simply adopted the random hybrid beamformers or DFT based hybrid beamformers for DOA measurements \cite{2018FanD, 2018Zheng, shu_low-complexity_2018, 2012Liu}, which requires a large number of training pilots to guarantee good performance.  In this letter we investigate the DOA estimation and optimize the HBF to improve the performance based on the Cramér-Rao bound (CRB) analysis with the utilization of the prior information of the DOA range. Our contributions can be summarized as follows:
\begin{itemize}
\item Recognizing that many existing estimation algorithms, e.g., \cite{2017Fan, shu_low-complexity_2018}, perform closely to the CRB, we propose to optimize HBF aiming at minimizing the CRB. The simulation results verify that our optimized HBF can improve the performance of  existing DOA estimation algorithms.

\item As there is usually some prior information about the (rough) range of DOA, we elaborate how to utilize the prior information to specify the optimization objective for better performance. 

\item Due to the partially-connected (PC) HBF architecture and the implementation of the phase shifters, the feasible region of the CRB minimization problem is  non-convex, which complicates the solution. To tackle the highly non-convexity, we propose an efficient manifold optimization (MO) based algorithm with guaranteed convergence.
\end{itemize}

\emph{Notations:} Matrices and vectors are denoted by boldface capital and lower-case letters, respectively. $[\mathbf{a}]_i$ denotes the $i$-th entry of a vector. $[\mathbf{A}]_{ij}$ denotes the $(i,j)$-th entry of a matrix. $(\cdot)^*$, $(\cdot)^T$, and $(\cdot)^H$ denote the complex conjugate, transpose, complex conjugate transpose of a matrix or vector. $\mathrm{d}(\mathbf{A}$) denotes the differential of $\mathbf{A}$. $\mathrm{tr(\cdot)}$,  $\|.\|_F$, and $\mathrm{Re}(\cdot)$ denote the trace,  the Frobenius norm, and  of the real part of a matrix, respectively. $\mathrm{diag}(\mathbf{x})$ is a diagonal matrix with the entries of $\mathbf{x}$ on its main diagonal and $\mathrm{blkdiag}(\mathbf{X}_1,\dots,\mathbf{X}_n)$ denotes a block diagonal matrix whose diagonal components are $\mathbf{X}_1,\dots,\mathbf{X}_n$.  $\mathcal{CN}(\mathbf{0}, \mathbf{K})$ denotes  the  circularly symmetric complex Gaussian distribution with zero mean and  covariance matrix $\mathbf{K}$. $\otimes$ and $\odot$ denote the Kronecker product and the Hadamard product, respectively.

% ---------------------------------
% ---- Section II System Model ----
% ---------------------------------
\section{System Model}\label{sec:systemmodel}
%Consider an mmWave massive MIMO communication system operated in  time division duplex (TDD) mode. The base station (BS) equipped with  $N_\mathrm{BS}$  antennas serves the user equipment (UE) equipped with   $N_\mathrm{UE}$  antennas. The BS side  adopts the HBF technique with $N_\mathrm{RF}$ RF chains, noticed that $N_\mathrm{RF}\le N_\mathrm{BS}$.  To  further avoid high power consumption and hardware cost,  the more practical partially-connected (PC)  architecture \cite{yu_alternating_2016, shu_low-complexity_2018} is considered, i.e., each antenna only connects to one single phase shifter.

%As an initial work and for the easy of presentation, we elaborate the idea under a scenario of single path channel model for more intuitive derivations of CRB \cite{shu_low-complexity_2018}. Nevertheless, it can be regarded as the approximation of typical LoS-dominated mmWave channels due to the well-known sparsity \cite{gao_reliable_2017} and  we also reveal the its robustness and potential of extensions to multi-paths scenarios via simulations.  

Consider the DOA estimation in the uplink of a block-fading mmWave MIMO communication system, where a base station (BS) is equipped with a large number ($N_\mathrm{BS}$) of antennas and adopts the PC-HBF architecture to reduce the hardware cost, and a user equipment (UE) is equipped with a small number ($N_\mathrm{UE}$) of antennas and adopts the fully digital beamforming. Define the transmitted training sequence at the UE as $\mathbf{s}=[s_0,\ldots,s_{N-1}]^T$, where $N$ is the length and $|s_n|=1$. As the training sequence ${\mathbf s}$ are being beamformed by ${\mathbf v} \in {\mathbb C}^{N_\mathrm{UE}}$, the equivalent baseband received signal at the BS antenna array is given by \cite{JLee2016, Alkhatted2014}
\begin{equation}
    \mathbf{r}_n = \mathbf{H}\mathbf{v} s_n + \mathbf{z}_n, \quad \mathrm{for}\;\; n=0,\ldots, N-1,
\end{equation}
where $\mathbf{z}_n$ denotes the additive Gaussian noise vector  with $\mathbf{z}_n \sim \mathcal{CN}(\mathbf{0}, \sigma^2\mathbf{I}_{N_\mathrm{BS}})$ and $\sigma^2$ represents the noise variance, the transmit power is represented by $P = \|\mathbf{v}\|^2$. $\mathbf{H}$ is the mmWave MIMO channel matrix and assumed  unchanged during the whole training process. Normally $\mathbf{H}$ can be characterized by the geometry-based channel model as follows
\begin{equation}\label{eqn:channel-model}
    \mathbf{H}= \sqrt{\frac{N_\mathrm{BS}N_\mathrm{UE}}{L}} \sum_{l=0}^{L-1} \alpha_l \mathbf{a}_{\mathrm{BS},l}(\theta_l, \phi_l) \mathbf{a}_{\mathrm{UE},l}^{H}(\psi_l, \gamma_l),
\end{equation}
where $L$ is the number of propagation paths and $l=0$ denotes the line-of-sight (LoS) link which has the strongest gain. Furthermore, $\alpha_l$ is the complex path gain of the $l$-th path, $\mathbf{a}_{\mathrm{BS},l}$ and $\mathbf{a}_{\mathrm{UE},l}$ represent the antenna array response vectors of the BS and the UE, respectively. $\theta_l (\phi_l)$ denotes the associated azimuth (elevation) angle of arrival, and $\psi_l (\gamma_l)$ denotes the associated azimuth (elevation) angle of departure, respectively. Given that a uniform planar array (UPA) of $P \times Q$ elements is deployed at the BS, the array response is  \cite{yu_alternating_2016,ayach_spatially_2014}
\begin{equation}
\label{USPA}
    \mathbf{a}_{\mathrm{BS},l}(\theta_l, \phi_l) = \mathbf{a}_y(\theta_l, \phi_l) \otimes \mathbf{a}_z(\phi_l),
\end{equation}
where $\mathbf{a}_{y}(\theta_l, \phi_l) = \frac{1}{\sqrt{Q}}[1, e^{\mathrm{j}\pi\sin\theta_l\sin\phi_l}, \dots, e^{\mathrm{j}\pi(Q-1)\sin\theta_l\sin\phi_l}]^T$ and $\mathbf{a}_{z}(\phi_l) = \frac{1}{\sqrt{P}}[1, e^{\mathrm{j}\pi\cos\phi_l}, \dots, e^{\mathrm{j}\pi(P-1)\cos\phi_l}]^T$.
Denote the hybrid combiner at the BS as $\mathbf{W}_n=\mathbf{W}_{\mathrm{RF},n}\mathbf{W}_{\mathrm{BB},n}\in \mathbb{C}^{N_\mathrm{BS}\times N_\mathrm{RF}}$, where $\mathbf{W}_{\mathrm{RF},n}\in \mathbb{C}^{N_\mathrm{BS}\times N_\mathrm{RF}}$ denotes the analog combiner and $\mathbf{W}_{\mathrm{BB},n}\in \mathbb{C}^{N_\mathrm{RF}\times N_\mathrm{RF}}$ denotes the digital baseband one. Then, the combined signal at time instance $n$  can be represented as 
\begin{equation}
    \mathbf{y}_n = \mathbf{W}_n^H \mathbf{r}_n = \mathbf{W}_n^H\mathbf{H}\mathbf{v} s_n + \mathbf{W}_n^H \mathbf{z}_n.
\end{equation}
It should be mentioned that due to the implementation of phase shifters in the PC-HBF architecture, $\mathbf{W}_{\mathrm{RF},n} = \mathrm{blkdiag}(\mathbf{w}_{n,1}, \dots, \mathbf{w}_{n, N_\mathrm{RF}})$, where $\mathbf{w}_{n, m}$, for $m=1,\dots, N_\mathrm{RF}$, is an $\frac{N_\mathrm{BS}}{N_\mathrm{RF}}\times 1$ column vector with its elements having a unit modulus, i.e., $|[\mathbf{w}_{n,m}]_{i}| = 1$, $\forall n,m,i$ \cite{yu_alternating_2016, shu_low-complexity_2018}.

\section{CRB Analysis and Problem Formulation}\label{sec:ML and CRLB}
Inspired by the fact that the existing DOA estimation algorithm can closely approach the CRB \cite{shu_low-complexity_2018, 2017Fan, 2012Liu}, we propose to optimize the hybrid combiner with the objective of minimizing the CRB. In this section we first analyze the CRB with HBF, and then formulate the HBF optimization problem for DOA estimation.

To simplify the analysis, we recognize that the downlink mmWave transmission is usually dominated by the LoS path due to its much higher gain compared with the none LoS (NLoS) paths \cite{gao_reliable_2017}. Thus, in the following derivation, we focus on the estimation of the DOA of the LoS path and ignore the effect of the NLoS paths. However, such effect will be considered in the simulation. 

%To simplify the analysis, we assume that the downlink transmission is usually dominated by the LoS link due to its much higher gain compared with the NLoS links, typically, $\alpha_1 = 10^{0.5} \alpha_i$ for $i=2\dots, P$ \cite{gao_reliable_2017}. Besides, the DOA estimation of the LoS path can also inform the BS of the location of UE, which is necessary for high-speed communications. In this paper, we focus on the estimation of the DOA of the LoS path.

\subsection{CRB Analysis}\label{subsec:crlb}
 As the training sequence is known at both the BS and the UE, we have 
\begin{equation}\label{eqn:y_tilde_n}
\tilde{\mathbf{y}}_n = \mathbf{y}_n s_n^* = \beta\mathbf{W}_n^H\mathbf{a}_{\mathrm{BS}}(\theta, \phi) + \mathbf{W}_n^H\tilde{\mathbf{z}}_n,
\end{equation}
where the subscript $(.)_{0}$ of the channel parameters of the LoS path is omitted for simplicity, $\beta = \alpha\mathbf{a}_{\mathrm{UE}}^H(\psi, \gamma)\mathbf{v}$, and $\tilde{\mathbf{z}}_n=s_n^*\mathbf{z}_n$ has the same distribution as $\mathbf{z}_n$. By collecting $\tilde{\mathbf{y}}_n$ for $n = 0, \dots, N-1$,  we have
\begin{equation}\label{eqn:y_tilde_vector}
   \tilde{\mathbf{y}} = \beta\mathbf{W}^H\mathbf{a}_{\mathrm{BS}}(\theta, \phi) + \widehat{\mathbf{W}}^H\tilde{\mathbf{z}},
\end{equation}
where $\tilde{\mathbf{y}} = [\tilde{\mathbf{y}}^T_0, \dots, \tilde{\mathbf{y}}^T_{N-1}]^T$, $\mathbf{W} = [\mathbf{W}_0, \dots, \mathbf{W}_{N-1}]$,  $\widehat{\mathbf{W}} = \mathrm{blkdiag}(\mathbf{W}_0, \dots,\mathbf{W}_{N-1})$, and $\tilde{\mathbf{z}}= [\tilde{\mathbf{z}}_0^T, \dots, \tilde{\mathbf{z}}_{N-1}^T]^T$. To derive the CRB, first define $\boldsymbol{\eta} \triangleq [{\theta}, \phi, \mathrm{Re}\{{\beta}\}, \mathrm{Im}\{{\beta}\}]$ as the vector containing the parameters to be estimated. As $\tilde{\mathbf{y}}\sim\mathcal{CN}\left(\beta\mathbf{W}^H\mathbf{a}_{\mathrm{BS}}(\theta, \phi), \sigma^2\widehat{\mathbf{W}}^H\widehat{\mathbf{W}}\right)$, according to \cite{2020CRLB}, the Fisher information matrix (FIR) can be derived as follows 
%\begin{equation}\label{eqn:yCN}
%    \mathbf{y} \sim \mathcal{CN}(\beta\mathbf{W}^H\mathbf{a}_{\mathrm{BS}}(\theta, \phi), \sigma^2\hat{\mathbf{W}}^H\hat{\mathbf{W}})).
%\end{equation}
\begin{equation}\label{eqn:Fisher}
    \mathbf{F}=\frac{2}{\sigma^{2}}\operatorname{Re}\{\mathbf{A}^H\mathbf{W}(\widehat{\mathbf{W}}^H\widehat{\mathbf{W}})^{-1}\mathbf{W}^H\mathbf{A}\},
\end{equation}
where $\mathbf{A} = \frac{\partial (\beta\mathbf{a}_{\mathrm{BS}}(\theta, \phi))}{\partial \boldsymbol{\eta}}$ is an $N_\mathrm{BS}\times 4$ matrix and is given by
\begin{equation}\label{eqn:compute_Fisher_A}
  \mathbf{A} = [\beta\mathbf{a}_1, \beta\mathbf{a}_2, \mathbf{a}_{\mathrm{BS}}(\theta, \phi), \mathrm{j}\mathbf{a}_{\mathrm{BS}} (\theta,\phi)],
\end{equation}
where 
\begin{equation}
\begin{split}
    &\mathbf{a}_1 =  \left({\mathrm{j}\pi\cos\theta\sin\phi}[0, 1\dots, Q-1]^T\odot \mathbf{a}_y(\theta,\phi)\right)\otimes \mathbf{a}_z(\phi) \\
    &\mathbf{a}_2 =  \left({\mathrm{j}\pi\sin\theta\cos\phi}[0, 1\dots, Q-1]^T\odot \mathbf{a}_y(\theta,\phi)\right)\otimes \mathbf{a}_z(\phi)\\
    &\;\;\;\;\;\;\;+\mathbf{a}_y(\theta,\phi)\otimes \left(-{\mathrm{j}\pi\sin\phi}[0, 1, \dots, P-1]^T\odot\mathbf{a}_z(\phi)\right).\nonumber
    \end{split}
\end{equation}
Recalling that $\mathbf{W}_n=\mathbf{W}_{\mathrm{RF},n}\mathbf{W}_{\mathrm{BB},n}$, we have
\begin{equation}\label{eqn:W-concatenation}
    \mathbf{W} = \mathbf{W}_\mathrm{RF}\mathbf{W}_\mathrm{BB}, \quad  \widehat{\mathbf{W}} = \widehat{\mathbf{W}}_\mathrm{RF}\mathbf{W}_\mathrm{BB},
\end{equation}
where $\mathbf{W}_\mathrm{RF}=[\mathbf{W}_{\mathrm{RF}, 0}, \dots, \mathbf{W}_{\mathrm{RF}, N-1}] \in \mathbb{C}^{N_\mathrm{BS}\times N_\mathrm{RF}N}$, $\mathbf{W}_\mathrm{BB}= \mathrm{blkdiag}(\mathbf{W}_{\mathrm{BB}, 0}, \dots, \mathbf{W}_{\mathrm{BB}, N-1}) \in \mathbb{C}^{N_\mathrm{RF}N\times N_\mathrm{RF}N}$, and  $\widehat{\mathbf{W}}_\mathrm{RF}= \mathrm{blkdiag}(\mathbf{W}_{\mathrm{RF}, 0}, \dots, \mathbf{W}_{\mathrm{RF}, N-1}) \in \mathbb{C}^{N_\mathrm{BS}N\times N_\mathrm{RF}N}$. By substituting \eqref{eqn:W-concatenation} into \eqref{eqn:Fisher}, we find that $\mathbf{F}$ can be simplified as 
\begin{equation}\label{eqn:simple-F}
     \mathbf{F}{=}\frac{2N_\mathrm{RF}}{\sigma^{2}N_\mathrm{BS}}\operatorname{Re}\{\mathbf{A}^H\mathbf{W}_\mathrm{RF}\mathbf{W}_\mathrm{RF}^H\mathbf{A}\}, 
\end{equation}
which follows from the fact that $\mathbf{W}_\mathrm{BB}$ is an invertable matrix and $\widehat{\mathbf{W}}_\mathrm{RF}^H\widehat{\mathbf{W}}_\mathrm{RF}=\frac{N_\mathrm{BS}}{N_\mathrm{RF}}\mathbf{I}$. Then, the CRB matrix  $\mathbf{C} = \mathbf{F}^{-1}$, where the diagonal elements reveal the minimum variances of the associated estimates. As we focus on the estimation of $\theta$ and $\phi$, we are interested in the left-top $2\times 2$ sub-matrix of $\mathbf{C}$, which is denoted by $\mathbf{C}_{11}$. By rewriting $\mathbf{F}$ as a block matrix, we have %the following expression for $\mathbf{C}_{11}$ based on the inversion formula of block matrix
\begin{equation}\label{eqn:BarC}
    \mathbf{F} = \left[\begin{array}{ll}
\mathbf{F}_{11} & \mathbf{F}_{12} \\
\mathbf{F}_{21} & \mathbf{F}_{22}
\end{array}\right], \;
    \mathbf{C}_{11} =  (\mathbf{F}_{11} - \mathbf{F}_{12}\mathbf{F}_{22}^{-1}\mathbf{F}_{21})^{-1},
\end{equation}
where  $\mathbf{F}_{mn}= \mathrm{Re}\{{\mathbf{A}}^H_m\mathbf{W}_\mathrm{RF}\mathbf{W}_\mathrm{RF}^H{\mathbf{A}}_n\}$ for $m,n=\{1,2\}$, and ${\mathbf{A}}_1$ and ${\mathbf{A}}_2$  are defined as the sub-matrices containing the first two columns and the last two columns of $\mathbf{A}$ in (\ref{eqn:compute_Fisher_A}), respectively.

%an optimization problem can be formulated to obtain a $\mathbf{W}_{\mathrm{RF}, \mathrm{opt}}$ which achieves a lower average CRB over that range. 
%$\mathbf{F}_{11}= \mathrm{Re}\{{\mathbf{A}}^H_1\mathbf{W}_\mathrm{RF}\mathbf{W}_\mathrm{RF}^H{\mathbf{A}}_1\}$, $\mathbf{F}_{12}=  \mathrm{Re}\{{\mathbf{A}}^H_1\mathbf{W}_\mathrm{RF}\mathbf{W}_\mathrm{RF}^H{\mathbf{A}}_2\}$, $\mathbf{F}_{21}= \mathrm{Re}\{{\mathbf{A}}^H_2\mathbf{W}_\mathrm{RF}\mathbf{W}_\mathrm{RF}^H{\mathbf{A}}_1\}$, and $\mathbf{F}_{22}= \mathrm{Re}\{{\mathbf{A}}^H_2\mathbf{W}_\mathrm{RF}\mathbf{W}_\mathrm{RF}^H{\mathbf{A}}_2\}$ are all $2\times 2$ matrices, where ${\mathbf{A}}_1$ and ${\mathbf{A}}_2$ are defined as the sub-matrices containing the first two columns and the last two columns of $\mathbf{A}$ in (\ref{eqn:compute_Fisher_A}), respectively.

\subsection{Problem Formulation}\label{subsec:formulation}
%From \eqref{eqn:y_tilde_vector}, we can see that as the baseband signal for DOA estimation is the output of the hybrid combiner $\mathbf{W}$, the estimation performance is closely related to the design of $\mathbf{W}$. 
From \eqref{eqn:BarC}, we see that the CRB is a function of $\mathbf{W}_\mathrm{RF}$. Thus, one can improve the estimation performance via optimizing $\mathbf{W}_\mathrm{RF}$ to minimize the CRB. That is, to solve
\begin{equation}\label{eqn:opt-W1}
    \mathbf{W}_{\mathrm{RF}, \mathrm{opt}} =  \arg\min_{\mathbf{W}_\mathrm{RF}} \mathrm{tr}({\mathbf{C}_{11}}). 
\end{equation}
The CRB, however, is associated with the unknown $\beta$ and DOA, and thus cannot be directly used as the objective function. Nevertheless, we first propose the following lemma:
\begin{lemma} \label{lemma1}
The solution of (\ref{eqn:opt-W1}) is independent of $\beta$.
\end{lemma}
\textit{Proof}: According to \eqref{eqn:Fisher} and the defination of $\mathbf{F}_{mn}$,  both $\mathbf{F}_{11}$ and $\mathbf{F}_{12}\mathbf{F}_{22}^{-1}\mathbf{F}_{21}$ in \eqref{eqn:BarC} are of  form  $\beta^2 f(\theta,\phi, \mathbf{W}_\mathrm{RF})$, where $f(\theta,\phi, \mathbf{W}_\mathrm{RF})$ is a function not related to $\beta$. Thus, $\mathrm{tr}({\mathbf{C}_{11}})$ is of form $\beta^{-2} f(\theta,\phi, \mathbf{W}_\mathrm{RF})$. As $\beta$ is not a function of $\mathbf{W}$ according to its definition, the solution of (\ref{eqn:opt-W1}) is not a function of $\beta$. This completes the proof. $\hfill\blacksquare$ 

According to  Lemma 1, $\beta$ can be set to $1$ in the following derivation without loss of generality. Further, if the range of the DOA to be estimated is known a priori, we can then optimize $\mathbf{W}_\mathrm{RF}$ to minimize the average CRB over that range. Denoting the prior ranges of the azimuth and elevation angles as $[\theta_\mathrm{b}, \theta_\mathrm{u}]$ and $[\phi_\mathrm{b}, \phi_\mathrm{u}]$, respectively, we uniformly sample them as follows
\begin{equation}\label{eqn:theta-k}
\begin{split}
\theta_j &= \theta_\mathrm{b} + \frac{(j-1)}{J}(\theta_\mathrm{u} - \theta_\mathrm{b}), \quad \mathrm{for}\; j = 1, \dots, J,\\
\phi_k &= \phi_\mathrm{b} + \frac{(k-1)}{K}(\phi_\mathrm{u} - \phi_\mathrm{b}), \quad \mathrm{for}\; k = 1, \dots, K.
\end{split}
\end{equation}
Instead of solving the problem in \eqref{eqn:opt-W1}, we optimize  $\mathbf{W}_{\mathrm{RF}}$ aiming at minimizing the average CRB over the sampled DOA range subject to the unit modulus constraints. That is, 
\begin{equation}\label{eqn:opt_prob-W_RF}
\begin{array}{cl}
\displaystyle{\minimize_{\mathbf{W}_\mathrm{RF}}} &   f=\sum\limits_{j=1}^J \sum\limits_{k=1}^K \mathrm{tr}(\mathbf{C}_{11}(\theta_j,\phi_k))\\
\mathrm{subject \; to} &  |[\mathbf{w}_{n,m}]_{i}| = 1, \quad\forall n,m,i. 
\end{array}
\end{equation}
where $\mathbf{C}_{11}(\theta_j,\phi_k)$ corresponds to the one by replacing $\theta$ and $\phi$ with $\theta_j$ and $\phi_k$ in \eqref{eqn:BarC}.
%\begin{equation}\label{eqn:W-discrete-angles}
%    \mathbf{W}_{\mathrm{RF}, \mathrm{opt}} = \arg\min_{\mathbf{W}_\mathrm{RF}}\sum_{j=1}^{J}\sum_{k=1}^K\mathrm{tr}({\mathbf{C}}_{11}(\theta_j,\phi_k)),
%\end{equation}
%where $\mathbf{C}_{11}(\theta_j,\phi_k)$ corresponds to the one by replacing $\theta$ and $\phi$ with $\theta_j$ and $\phi_k$ in \eqref{eqn:BarC}. Finally, consider the constant modulus constraint introduced by the analog combiner $\mathbf{W}_{\mathrm{RF},n}$,  the optimization problem can be given by

%substituting each combination of $\theta_j$ and $\phi_k$ into (\ref{eqn:compute_Fisher}) results in an individual $\Bar{\mathbf{C}}$ associated with $\mathbf{W}$. As the optimized $\mathbf{W}_*$ is supposed to well estimate the arbitrary angles in the ranges rather than one pair of particular $\theta$ and $\phi$, we intuitively propose the following optimization problem to seek for a general solution  
%\begin{equation}\label{eqn:W*}
%    \mathbf{W}_* = \arg\min_{\mathbf{W}}\sum_{j=1}^{J}\sum_{k=1}^K\mathrm{tr}(\Bar{\mathbf{C}}_{j,k}),
%\end{equation}
%where $\Bar{\mathbf{C}}_{j,k}\triangleq \Bar{\mathbf{C}}|_{\theta=\theta_j, \phi = \phi_k}$ are defined for notational brevity. The objective of \eqref{eqn:W*} with large $J$ and $K$ can be regarded as the approximation of the performance upper bound averaged over all possible DOAs.  

% =========== Section IV ============
\section{Manifold Optimization Algorithm}\label{sec:alg}
% ===================================

%Although the objective function is differentiable and the traditional gradient-descend (GD) algorithm can be applied to find a local minimizer of (\ref{eqn:opt_prob-W_RF}), the solution is not guaranteed to be feasible.
It appears difficult to solve (\ref{eqn:opt_prob-W_RF}) because of not only the  complicated objective function, but also the highly non-convex feasible set. Specifically, the analog beamformer $\mathbf{W}_{\mathrm{RF},n}$ has a block diagonal structure and only the non-zero elements in the block matrices need to be optimized. Furthermore, they should satisfy the unit modulus constraint. To tackle these difficulties, we first introduce a sparse mask matrix $\mathbf{P}\in\mathbb{C}^{N_\mathrm{BS}\times N_\mathrm{RF}N}$ as 
\begin{equation}
    \mathbf{P}_{ij}=\Big\{
\begin{array}{ll}
1 & [\mathbf{W}_\mathrm{RF}]_{ij}  \ne   0 \\
0 & [\mathbf{W}_\mathrm{RF}]_{ij}  =   0,
\end{array}
\end{equation}
then we can rewrite the analog beamformer in a form of  $\mathbf{W}_\mathrm{RF}= \mathbf{P} \odot {\widetilde{\mathbf{W}}}_\mathrm{RF}$, where  ${\widetilde{\mathbf{W}}}_\mathrm{RF}\in\mathbb{C}^{N_\mathrm{BS}\times N_\mathrm{RF}N}$ is an  auxiliary matrix variable without the block diagonal matrix constraint and all of its elements should satisfy the unit modulus constraints. Thus,  the feasible set of ${\widetilde{\mathbf{W}}}_\mathrm{RF}$ is essentially a typical Riemannian manifold \cite{yu_alternating_2016,lin_hybrid_2019}, i.e., $\mathcal{X} = \{\mathbf{X}\in \mathbb{C}^{N_\mathrm{BS}\times  N_\mathrm{RF}N}:\left|\left[\mathbf{X}\right]_{i j}\right|=1, \forall i, j\}.$
Therefore, to minimize $f$ with respect to ${\widetilde{\mathbf{W}}}_\mathrm{RF}$ (instead of ${\mathbf{W}}_\mathrm{RF}$) becomes a Riemannian optimization problem, which has been studied in   \cite{yu_alternating_2016, lin_hybrid_2019}.% and abundant optimization tools in the Euclidean space (e.g., the GD and trust-region methods) can be transplanted for solution.  

In this letter, we propose to extend the gradient-descend (GD) algorithm to minimize the objective in (\ref{eqn:opt_prob-W_RF}) over the Riemannian manifold. The basic idea is that in the $i$-th iteration, we first update the optimization variable $\widetilde{\mathbf{W}}_\mathrm{RF}^{(i)}$ along the opposite direction of the \textit{Riemannian gradient} to achieve a local minimizer on its \textit{tangent space}, where the tangent space is a linear space composed of all the vectors that tangentially pass through $\widetilde{\mathbf{W}}_\mathrm{RF}^{(i)}$, and the Riemannian gradient is the projection of the  conjugate Euclidean gradient $\nabla f({\widetilde{\mathbf{W}}}_\mathrm{RF}^{(i)})$ onto the tangent space. Subsequently, we retract the minimizer on the tangent space back to the manifold to obtain $\widetilde{\mathbf{W}}_\mathrm{RF}^{(i+1)}$ as the finish of the iteration.

However, the application of manifold optimization is not straightforward and the conjugate Euclidean  gradient needs to be derived first. Based on the differential rule $\mathrm{d}(\mathbf{X}^{-1})=-\mathbf{X}^{-1}\mathrm{d}(\mathbf{X}^{-1})\mathbf{X}^{-1}$, we have from \eqref{eqn:BarC}
\begin{equation}\label{derive-equs2}
\begin{split}
    &\mathrm{d}\left(\mathrm{tr}(\mathbf{C}_{11})\right)=\mathrm{tr}(\mathbf{T}\mathrm{d}(\mathbf{F}_{11})) - \mathrm{tr}(\mathbf{T}\mathrm{d}(\mathbf{F}_{12})\mathbf{F}_{22}^{-1}\mathbf{F}_{21})\\
    &\;\;\;\;+ \mathrm{tr}(\mathbf{T}\mathbf{F}_{12}\mathbf{F}_{22}^{-1}\mathrm{d}(\mathbf{F}_{22})\mathbf{F}_{22}^{-1}\mathbf{F}_{21}) - \mathrm{tr}(\mathbf{T}\mathbf{F}_{12}\mathbf{F}_{22}^{-1}\mathrm{d}(\mathbf{F}_{21})),
\end{split}
\end{equation}
where $\mathbf{T}= -(\mathbf{F}_{11} - \mathbf{F}_{12}\mathbf{F}_{22}^{-1}\mathbf{F}_{21})^{-2}$. We can further obtain from \eqref{eqn:simple-F} that \footnote{According to the differential rule \cite{2020ComplexDerive}, the term $\mathbf{W}_\mathrm{RF}$ is regarded as a constant matrix during the derivation of the conjugate gradient.}
\begin{equation}\label{derive-equs}
    \mathrm{d}(\mathbf{F}_{mn}) = \gamma\left({\mathbf{A}}^H_m\mathbf{W}_\mathrm{RF}
    \mathrm{d}(\mathbf{W}_\mathrm{RF}^H){\mathbf{A}}_n + {\mathbf{A}}^T_m\mathrm{d}(\mathbf{W}_\mathrm{RF}^*)\mathbf{W}_\mathrm{RF}^T{\mathbf{A}}_n^*\right),\nonumber
\end{equation}
where $\gamma \triangleq \frac{N_\mathrm{RF}}{\sigma^2N_\mathrm{BS}}$.
Substituting them into \eqref{derive-equs2} and using the fact that $\mathrm{d}(\mathbf{W}_\mathrm{RF}) = \mathbf{P}\odot\mathrm{d}(\widetilde{\mathbf{W}}_\mathrm{RF})$ and $\mathrm{tr}\left(\mathbf{A}(\mathbf{B}\odot\mathbf{C})\right) = \mathrm{tr}\left((\mathbf{A}\odot\mathbf{B}^T)\mathbf{C}\right)$  for arbitrary matrices $\mathbf{A}$, $\mathbf{B}$ and $\mathbf{C}$, we have
\begin{equation}\label{derive-equs3}
\begin{split}
      \mathrm{d}\left(\mathrm{tr}(\mathbf{C}_{11})\right) &=\gamma \mathrm{tr}\bigg(\bigg(\big({\mathbf{A}}_1\mathbf{T}{\mathbf{A}}_1^H +  {\mathbf{A}}_2\mathbf{F}_{22}^{-1}\mathbf{F}_{21}\mathbf{T}\mathbf{F}_{12}\mathbf{F}_{22}^{-1}{\mathbf{A}}_2^H - \\ &2\mathrm{Re}\{{\mathbf{A}}_2\mathbf{F}_{22}^{-1}\mathbf{F}_{21}\mathbf{T}{\mathbf{A}}_1^H\}\big){{\mathbf{W}}}_\mathrm{RF}\odot \mathbf{P}\bigg)\mathrm{d}(\widetilde{\mathbf{W}}_\mathrm{RF}^H)\bigg). 
\end{split}
\end{equation}
%relationship between the conjugate gradient and the differential, i.e.,
According to  that $\mathrm{d}(f({\widetilde{\mathbf{W}}}_\mathrm{RF})) = \mathrm{tr}(\nabla f({\widetilde{\mathbf{W}}}_\mathrm{RF})\mathrm{d}({\widetilde{\mathbf{W}}}_\mathrm{RF}^H))$, we obtain the Euclidean gradient from \eqref{derive-equs3}
\begin{equation}\label{eqn:compute-g1}
    \nabla f({\widetilde{\mathbf{W}}}_\mathrm{RF})=\gamma\sum_{j=1}^J\sum_{k=1}^K\left(\left(\mathbf{J}_{jk} + \mathbf{K}_{jk} - 2\mathrm{Re}\{\mathbf{Q}_{jk}\} \right){{\mathbf{W}}}_\mathrm{RF}\right)\odot \mathbf{P},
\end{equation}
where 
\begin{equation}\label{eqn:compute-g2}
    \begin{split}
      \mathbf{J}_{jk} &= ({\mathbf{A}}_1\mathbf{T}{\mathbf{A}}_1^H)|_{\theta=\theta_j, \phi = \phi_k},\\
     \mathbf{K}_{jk} &= ({\mathbf{A}}_2\mathbf{F}_{22}^{-1}\mathbf{F}_{21}\mathbf{T}\mathbf{F}_{12}\mathbf{F}_{22}^{-1}{\mathbf{A}}_2^H)|_{\theta=\theta_j, \phi = \phi_k},\\
     \mathbf{Q}_{jk} &= ({\mathbf{A}}_2\mathbf{F}_{22}^{-1}\mathbf{F}_{21}\mathbf{T}{\mathbf{A}}_1^H)|_{\theta=\theta_j, \phi = \phi_k}.
    \end{split}
\end{equation}
%More details can be referred to \cite{2020ComplexDerive}. 
The Riemannian gradient can be obtained by projecting the  Euclidean gradient onto the tangent space of ${\widetilde{\mathbf{W}}}_\mathrm{RF}$, i.e.,
\begin{equation}\label{eqn:Riemannian}
    \mathrm{grad}f\left({\widetilde{\mathbf{W}}}_\mathrm{RF}\right)=\nabla f\left({\widetilde{\mathbf{W}}}_\mathrm{RF}\right)-\mathrm{Re}\{ \nabla f\left({\widetilde{\mathbf{W}}}_\mathrm{RF}\right)\odot {\widetilde{\mathbf{W}}}_\mathrm{RF}^*\}\odot {\widetilde{\mathbf{W}}}_\mathrm{RF}.
\end{equation}
With the derived Riemannian gradient,  ${\widetilde{\mathbf{W}}}_\mathrm{RF}$ in the $i$-th iteration is updated as follows
\begin{equation}\label{eqn:retract}
    [{\widetilde{\mathbf{W}}}_\mathrm{RF}^{(i)}]_{pq} = \frac{[{\widetilde{\mathbf{W}}}_\mathrm{RF}^{(i-1)} + \alpha^{(i)} \mathbf{D}^{(i)}]_{pq}}{|[{\widetilde{\mathbf{W}}}_\mathrm{RF}^{(i-1)} + \alpha^{(i)}\mathbf{D}^{(i)}]_{pq}|},
\end{equation}
where $\mathbf{D}^{(i)}=-\mathrm{grad}f({\widetilde{\mathbf{W}}}_\mathrm{RF}^{(i-1)})$ and $\alpha^{(i)}$ denote the negative direction of Riemannian gradient and the Armijo backtracking step size, respectively. According to \cite{boumal2020intromanifolds, absil2009optimization},  ${\widetilde{\mathbf{W}}}_\mathrm{RF}$ is guaranteed to converge to a local minimum of $f({\widetilde{\mathbf{W}}}_\mathrm{RF})$ and satisfy the unit modulus constraints. The overall algorithm is summarized in Algorithm 1 and  termed as CRB-MO,  where $\epsilon$ is the convergence threshold. It is worth noting that as $\mathbf{W}$ can be optimized offline, it thus does not lead to any extra computational complexity in the real-time implementation.

\begin{algorithm}[t]
\label{alg:manifold}
	\caption{CRB-MO Algorithm}
	\begin{algorithmic}[1]
			\STATE Randomly initialize ${\widetilde{\mathbf{W}}}_\mathrm{RF}^{(0)}$ and set $i=0$;
\REPEAT	
	\STATE  Compute the Riemannian gradient $\mathrm{grad}f({\widetilde{\mathbf{W}}}_\mathrm{RF}^{(i)})$ according to \eqref{eqn:compute-g1} and \eqref{eqn:Riemannian};
	\STATE  Update ${\widetilde{\mathbf{W}}}_\mathrm{RF}^{(i+1)}$ according to \eqref{eqn:retract}; 
    \STATE  $i\leftarrow i+1$;

\UNTIL $f({\widetilde{\mathbf{W}}}_\mathrm{RF}^{(i-1)})-f({\widetilde{\mathbf{W}}}_\mathrm{RF}^{(i)})\le \epsilon$; 
                \STATE $\mathbf{W}_\mathrm{RF} = \mathbf{P} \odot {\widetilde{\mathbf{W}}}_\mathrm{RF}^{(i)}$;
	\end{algorithmic}
\end{algorithm}

\section{Simulation Results}\label{sec:simulation}
%In this section, we provide simulation results to evaluate the DOA estimation performance with the proposed CRB-MO HBF design algorithm. 
Throughout the simulations, we set $N_\mathrm{BS}=512$ ($P=16$, $Q=32$), $N_\mathrm{UE}=4$, $N_\mathrm{RF}=4$ and $N=4$. Without loss of generality, a typical maximum likelihood (ML) based algorithm in \cite{Li1993} is adopted for DOA estimation with different receive beamformers, i.e., the traditional random beamformers and the beamformers optimized via proposed CRB-MO algorithm with $J=K =180$ to guarantee sufficient angular resolutions. The mean square error (MSE) of the azimuth angle $\theta$ is adopted as the performance metric in the following figures, while the MSE of the elevation angle $\phi$ has been observed with similar result. The SNR is defined as $\frac{P}{\sigma^2}$. All the results were obtained from the average over $1000$ independent channel realizations.

\begin{figure}[t]
\centering
\begin{minipage}[t]{0.24\textwidth}
\centering
\includegraphics[height=4.8cm,width=4.8cm]{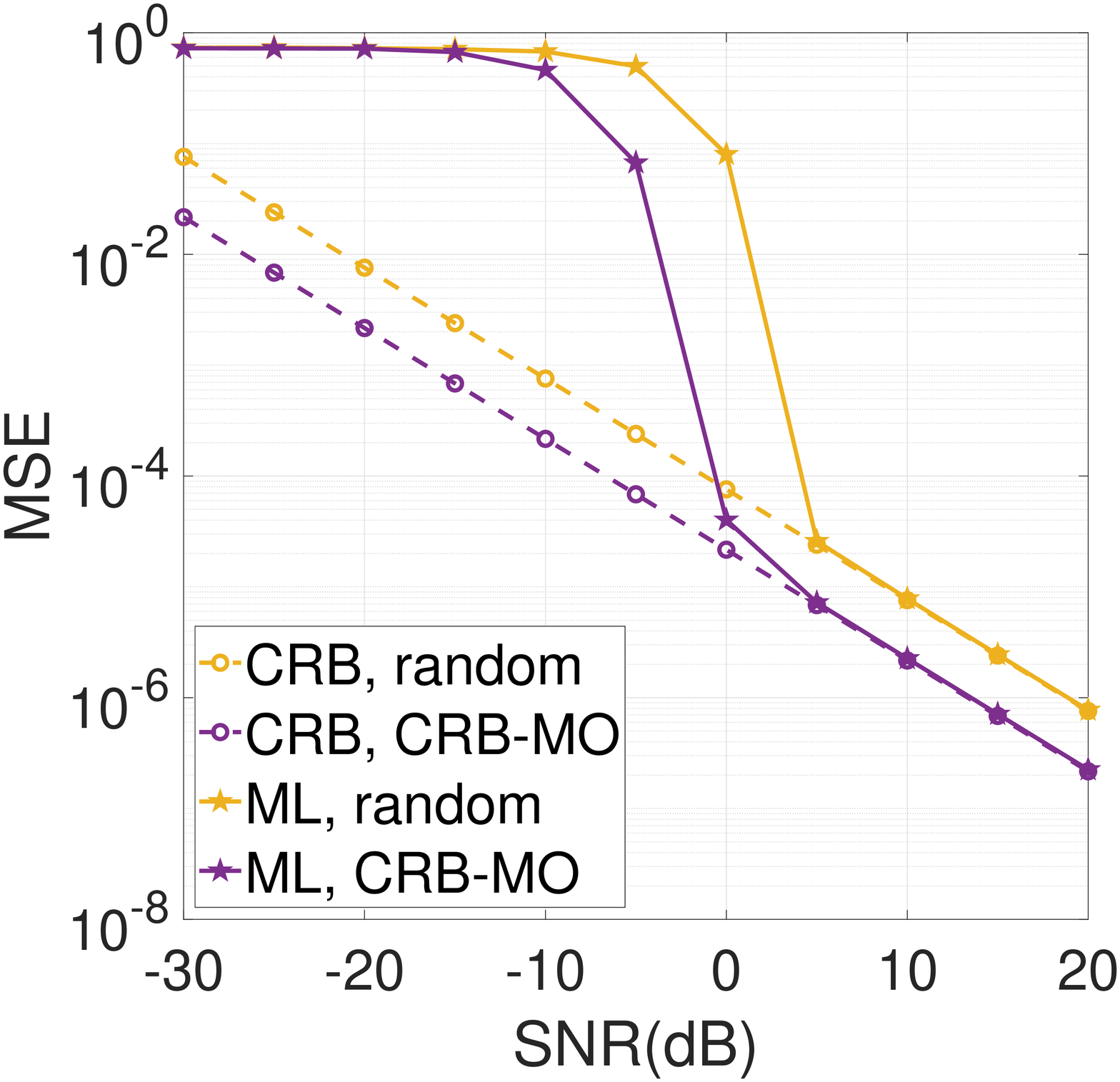}
\caption{MSE v.s. SNR for different HBF algorithms when $\theta\in [-\frac{\pi}{3}, \frac{\pi}{3}]$ and $\phi\in [\frac{5\pi}{12}, \frac{7\pi}{12}]$.}
\label{fig:MSEvsSNR_broad}
\end{minipage}
\begin{minipage}[t]{0.24\textwidth}
\centering
\includegraphics[height=4.8cm,width=4.8cm]{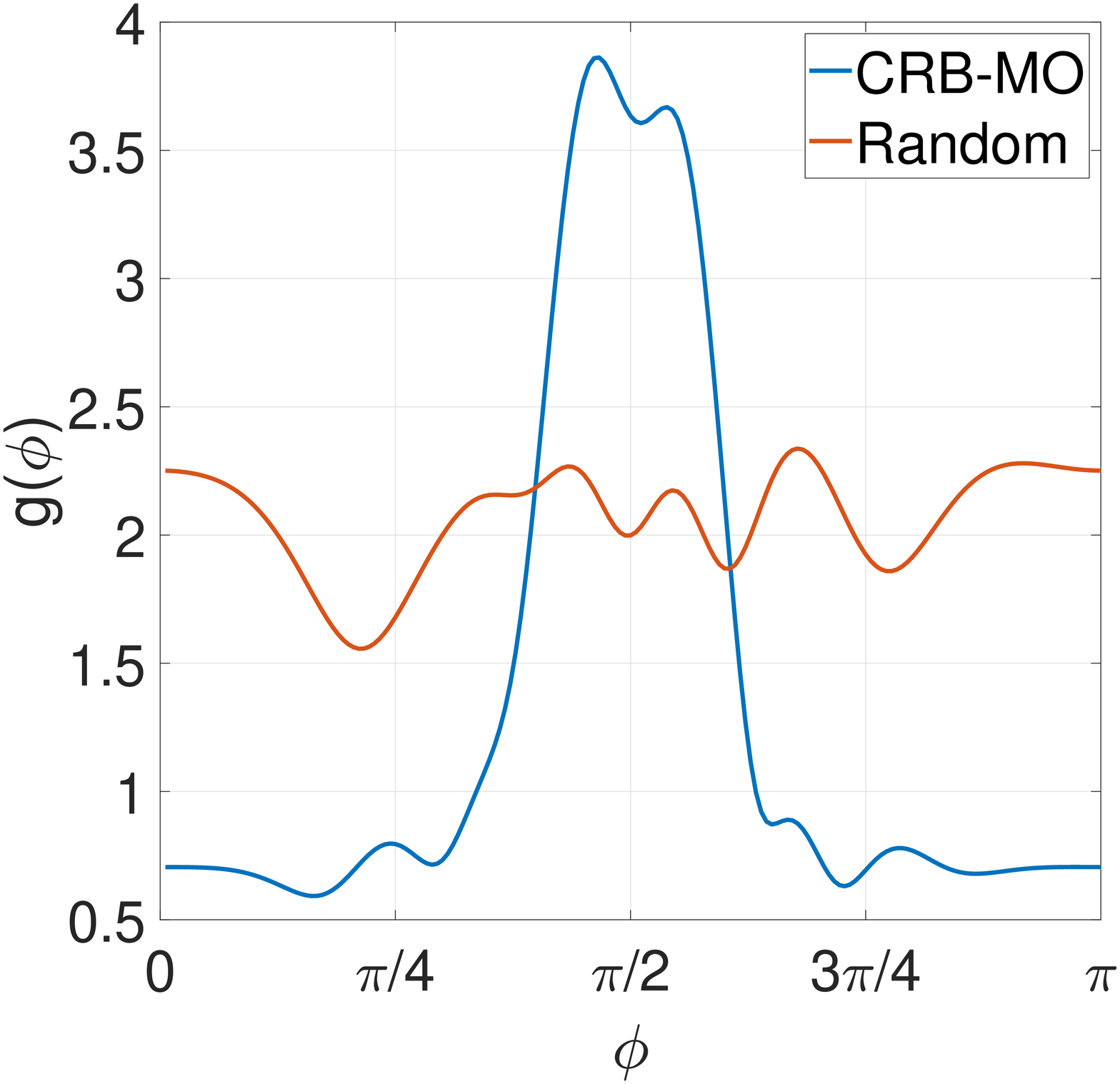}
\caption{Array power response $g(\phi)$ for different HBF algorithms.}
\label{fig:optimized_BF}
\end{minipage}
\end{figure}

Fig. \ref{fig:MSEvsSNR_broad} shows the average MSE as a function of SNR in a typical mmWave communication scenario with $\theta\in [-\frac{\pi}{3}, \frac{\pi}{3}]$, $\phi\in [\frac{5\pi}{12}, \frac{7\pi}{12}]$ and $\alpha\sim\mathcal{CN}(0,1)$. It is assumed that such DOA range is known a priori in the CRB-MO algorithm. We can see that with the optimized receive beamformers from the CRB-MO algorithm, the CRB is significantly improved by around $5\mathrm{dB}$ in the required SNR over that with randomly generated beamformers. Meanwhile, the ML DOA algorithm with the optimized beamformers achieves similar performance improvement and approaches the CRB.

To better explain the phenomenon in Fig. \ref{fig:MSEvsSNR_broad}, Fig. \ref{fig:optimized_BF} further depicts the array power response, which is defined as $g(\phi) = |\mathbf{W}\mathbf{a}_\mathrm{BS}(\theta, \phi)|$ with a fixed $\theta = \frac{\pi}{2}$, of the resulting beamformers of the two algorithms. We can see that the random beamformer exhibits a relatively flat power distribution in the whole angle domain. However, the proposed CRB-MO algorithm utilizes the prior information and generates a beam whose power is more concentrated on the specific DOA range. This provides some insight for the HBF design in the beam training stage for practical applications.

\begin{figure}[t]
\centering
\begin{minipage}[t]{0.24\textwidth}
\centering
\includegraphics[height=4.8cm,width=4.8cm]{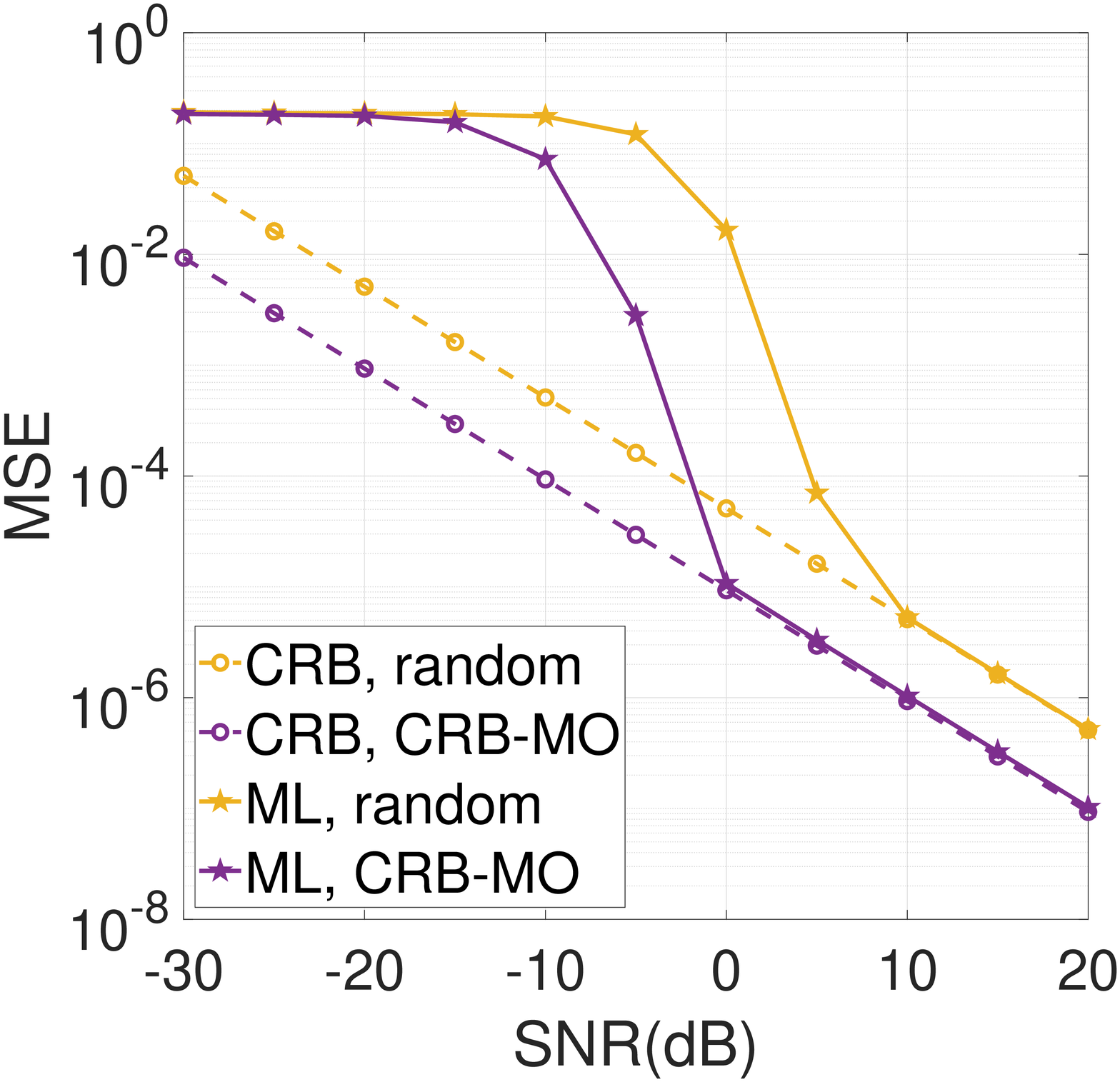}
\caption{MSE v.s. SNR for different HBF algorithms when $\theta\in [-\frac{\pi}{6}, \frac{\pi}{6}]$ and $\phi\in [\frac{5\pi}{12}, \frac{7\pi}{12}]$.}
\label{fig:MSEvsSNR_narrow}
\end{minipage}
\begin{minipage}[t]{0.24\textwidth}
\centering
\includegraphics[height=4.8cm,width=4.8cm]{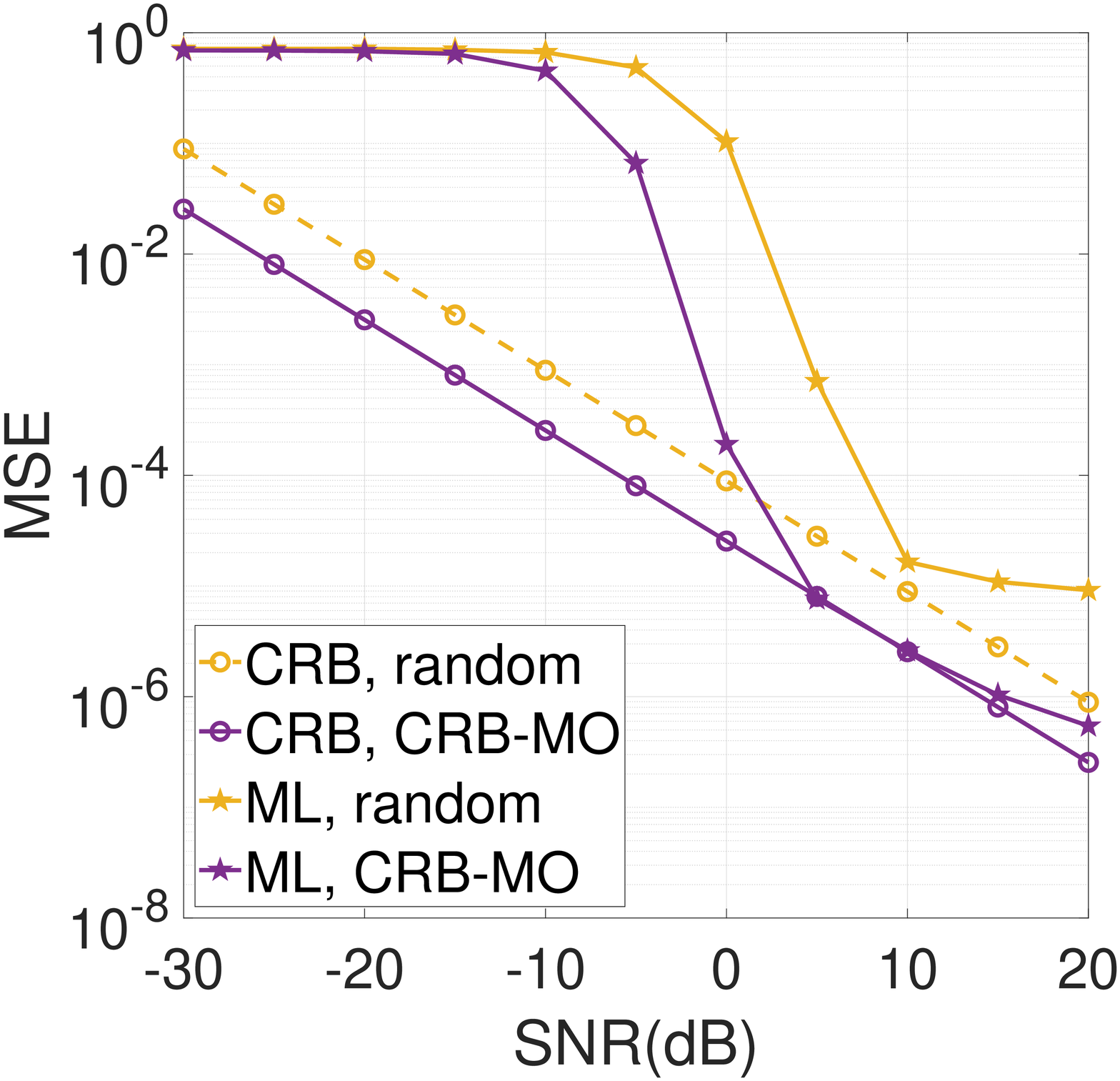}
\caption{Comparison of optimized HBF and random HBF under a multi-path channel model.}
\label{fig:MSEvsSNR_multipath}
\end{minipage}
\end{figure}

Fig. \ref{fig:MSEvsSNR_narrow} further demonstrates the comparison result with $\theta\in [-\frac{\pi}{6}, \frac{\pi}{6}]$ and $\phi\in [\frac{5\pi}{12}, \frac{7\pi}{12}]$, which can be regarded as a scenario in the warm boot stage where one may have more accurate information about the DOA range. Compared to the result in Fig. \ref{fig:MSEvsSNR_broad}, both algorithms achieve a lower MSE as the DOA range is narrowed. However, the CRB-MO algorithm achieves a higher gain due to the more specific prior information. This is because, as similar to that in Fig. \ref{fig:optimized_BF}, we observed a more concentrated power distribution with a narrower DOA range. 
%This phenomenon can also be explained by the insight of . Under the transmit power limitation, the narrower the range needs to be focused, the larger the array response of the hybrid beamformer results from the CRB-MO algorithm.

%which corresponds to higher estimation SNR and lower CRB.   

%\alpha_0\sim\mathcal{CN}(0,1)$ \alpha_1\sim\mathcal{CN}(0,10^{-0.5})$

Finally, Fig. \ref{fig:MSEvsSNR_multipath} depicts the estimation result in a typical two-path scenario, where the power of the NLoS path is $-5\text{dB}$ lower than that of the LoS path \cite{gao_reliable_2017}. The DOA ranges of the two paths are set as follows: $\theta_0\in[-\frac{\pi}{3}, \frac{\pi}{3}]$, $\phi_0\in [\frac{5\pi}{12}, \frac{7\pi}{12}]$, $\theta_1\in[-\frac{\pi}{2}, \frac{\pi}{2}]$ and $\phi_1\in [0, \pi]$. For the CRB-MO algorithm, we only utilize the prior information of the LoS path. The CRB curves correspond to the joint estimation of the two paths based on the received signal, while the ML curves correspond to the DOA estimation of only the LoS path by taking the NLoS interference as part of the noise. Thus, at high SNRs, the ML curves become flat. However, it can be seen that the CRB-MO algorithm still significantly outperforms the random algorithm in the multi-path scenario. Although in this letter we focus on the DOA estimation of the LoS path, the proposed CRB-MO algorithm can also be extended to the beamformer design for the joint DOA estimation of the multiple paths.

\section{Conclusion}\label{sec:conclusion}
This letter proposed an HBF design approach for improving the DOA estimation performance based on the CRB analysis. By exploring the a prior information of the DOA range, we formulated an HBF optimization problem aiming at minimizing the average CRB over the prior DOA range subject to the constraint on the PC analog beamformer, and solved it by applying MO with guaranteed convergence. Simulation results have demonstrated the substantial performance improvement of the proposed CRB-MO algorithm over the convectional random beamforming. 

%propose a MO-based algorithm for the optimization of the hybrid beamformers for DOA estimation. By minimizing the corresponding average CRBs, significant performance improvement is achieved. %This paper provides the design insights for the DOA estimation, and it is of great interests to extend for more complicated scenarios in our future work.

\newpage
\bibliography{LT}
\bibliographystyle{IEEEtran}
\end{document}